\documentclass[prd,floatfix,preprintnumbers,letterpaper]{revtex4}
\usepackage{amsmath,amssymb,graphicx}
\usepackage{hyperref}

\usepackage{xcolor}
\usepackage{graphicx}
\usepackage{amssymb}
\usepackage{multirow,bigdelim}

\begin{document}
\title{The gas depletion factor in galaxy clusters: implication from Atacama Cosmology Telescope Polarization experiment measurements}
\author{Xiaogang Zheng$^{1,2}$, Jing-Zhao Qi$^{2,3}$, Shuo Cao $^{2\ast}$, Tonghua Liu
$^{2}$, Marek Biesiada$^{2,4}$, Sylwia Miernik$^{4}$ and Zong-Hong
Zhu$^{1,2}$ }

\affiliation{1. School of Physics and Technology, Wuhan University, 430072, Wuhan, China;\\
2. Department of Astronomy, Beijing Normal University,
100875, Beijing, China; \emph{caoshuo@bnu.edu.cn}\\
3. Department of Physics, College of Sciences,
Northeastern University, Shenyang 110004, China;\\
4. Department of Astrophysics and Cosmology, Institute of Physics,
University of Silesia, 75 Pu{\l}ku Piechoty 1, 41-500 Chorz{\'o}w,
Poland}

\begin{abstract}
The gas depletion factor $\gamma(z)$, i.e., the average ratio of the
gas mass fraction to the cosmic mean baryon fraction of galaxy
clusters, plays a very important role in the cosmological
application of the gas mass fraction measurements. In this paper,
using the newest catalog of 182 galaxy clusters detected by the
Atacama Cosmology Telescope (ACT) Polarization experiment, we
investigate the possible redshift evolution of $\gamma(z)$ through a
new cosmology-independent method. The method is based on
non-parametric reconstruction using the measurements of Hubble
parameters from cosmic chronometers. Unlike hydrodynamical
simulations suggesting constant depletion factor, our results reveal
the trend of $\gamma(z)$ decreasing with redshift. This result is
supported by a parametric model fit as well as by calculations on
the reduced ACTPol sample and on the alternative sample of 91 SZ
clusters reported earlier in ACT compilation. Discussion of possible
systematic effects leaves an open question about validity of the
empirical relation $M_{tot}$-$f_{gas}$ obtained on very close
clusters. These results might pave the way to explore the hot gas
fraction within large radii of galaxy clusters as well as its
possible evolution with redshift, which should be studied further on
larger galaxy cluster samples in the upcoming X-ray/SZ cluster
surveys.
\end{abstract}

\maketitle

\section{Introduction}\label{sec:introduction}

As the largest gravitational bound systems in the Universe, galaxy
clusters provide a particularly rich source of information about the
morphology of our accelerating universe \citep{Allen11}. In
particular, X-ray measurements of thermal bremsstrahlung radiation
from the hot intracluster medium (ICM) \citep{Vikhlinin09, Landry13,
Mantz14} and the Sunyaev-Zel'dovich (SZ) effect due to inverse
Compton scattering of CMB photons by electrons inside the hot ICM
\citep{SZ1972} became powerful tests. Especially, the latter effect
provides us an excellent probe of cosmology
\citep{Benson13,Planck16a,Haan16} and the dynamical properties of
massive galaxy clusters
\citep{Hasselfield13,Bleem15,Planck16b,Hilton17}. Much efforts have
also been made to explore the sizes of galaxy clusters with the
combination of X-ray emission from the ICM and the SZ effect
\citep{Filippis05,Bonamente06}. On the other hand, considering the
matter budget, X-ray emitting hot gas constituting the ICM dominates
the baryonic mass. The ratio of this gas mass to the total mass
(also known as gas mass fraction) in massive clusters of galaxies,
is deemed to approximately match the mean universal baryon fraction
as $f_{gas}\propto\gamma(\Omega_b/\Omega_m)$ with a depletion factor
$\gamma$, where $\Omega_b$ and $\Omega_m$ are the cosmic baryon
density and the total matter density, respectively
\citep{Borgani11}. It is obvious that such probe provides a robust
method to constrain the cosmic matter density $\Omega_m$, in
combination with the constraints on $\Omega_b$ from cosmic microwave
background (CMB) or big bang nucleosynthesis (BBN) data. The idea of
using the $f_{gas}$ measurements in clusters as a cosmological probe
was initiated by \citet{White93} and then developed successfully to
test modern competing cosmologies \citep{Allen08,Ettori09,Cao14}, as
well as the gas density and temperature profiles of galaxy clusters
\citep{Cao11a,Cao16}.

However, some problems arise when one uses $f_{gas}$ observations as
a cosmological tool, for instance, the assumption that the gas mass
fraction evolves little or does not evolve at all. 
Therefore, the derived $f_{gas}$ values should be calibrated with
the baryon depletion factor $\gamma$, the ratio by which the baryon
fraction of galaxy clusters is depleted with respect to the
universal mean of baryon fraction \citep{Allen08}. This implies that
evolutionary behavior of the depletion factor may play a crucial
role for the efficiency of $f_{gas}$ test. In order to quantify the
gas content and its possible evolution, parametrized as
$\gamma(z)=\gamma_0(1+\gamma_1z)$, \citet{Battaglia13,Planelles13}
investigated the depletion of X-ray emitting gas relative to the
cosmic baryon fraction with hydrodynamic simulations of massive
galaxy clusters with $M_{500}>2\times10^{14}h^{-1}M_{\odot}$. As
usual, $M_{500}$ denotes the total mass within $R_{500}$, the radius
inside which the mass density is 500 times the critical density of
the universe. Their results suggested the depletion factor inside
$R_{500}$ quantified with $\gamma_{0}=0.85\pm0.03$ and
$\gamma_1=0.02\pm0.07$, which indicated that at $z<1$ the gas mass
fraction at intermediate to large cluster radii should have small
cluster-to-cluster scatter and should not evolve with redshift. The
first attempt of studying the depletion factor with observational
data was performed by \citet{Holanda17a}, who used the luminosity
distances from type Ia supernova \citep{Betoule14} to calibrate the
gas mass fraction inside the $(0.8-1.2)\times r_{2500}$ shell using
\textit{Chandra} X-ray measurements \citep{Mantz14}. More recently,
with the combination of gas mass fraction obtained from X-ray
measurements \citep{LaRoque06} and angular diameter distances from
SZ effect/X-ray measurements, \citet{Holanda18} found a mild
redshift evolution of the depletion factor.

Several factors should be taken into account, however, in order to
appropriately assess the possible redshift evolution of the gas
depletion factor. First of all, either negligible or significant
evolution of $\gamma$ parameter might be just a statistical artefact
produced by not particularly rich or deep observational data used in
these studies. This suggests that the increased depth and quality of
observational data set may result with more firm and robust
conclusions \citep{Cao14}. Secondly, accuracy of distance
determination may strongly influence the estimated value of the
$\gamma$ parameter. For instance, in the framework of the Planck's
best-fitted $\Lambda$CDM cosmology \citep{Planck15}, there appears
no redshift evolution of the depletion factor for the gas mass
fraction from X-ray measurements of several galaxy clusters
\citep{Holanda18}. In this context, collection of more complete
observational data concerning the gas mass fraction does play a
crucial role. In our paper we turn to the largest SZ cluster sample
derived from observations by the Atacama Cosmology Telescope
Polarization experiment (ACTPol) \citep{Hilton17}, which comprises a
catalog of 182 galaxy clusters covering the redshift range of
$0.1<z<1.4$. Our purpose is to probe possible evolution of the gas
depletion factor inside $R_{500}$. Moreover, compared with the
previous works using the luminosity distances from type Ia supernova
\citep{Holanda17a,Holanda18}, we will use instead, angular diameter
distances covering the cluster redshift range derived in a
cosmological-model-independent way from cosmic chronometers' $H(z)$
measurements using Gaussian processes (GP).

The idea of cosmological application of GP technique in general and
with respect to $H(z)$ data in particular, was first discussed in
\citet{Holsclaw10} and then extensively applied in more recent
papers to test the cosmological parameters
\citep{Seikel12,Cao17b,Cao18}, the distance-duality relation
\citep{Zhang14}, spatial curvature of the Universe
\citep{Wei16,Cao19,Qi19c}, and the speed of light at higher
redshifts \citep{Cao17a}. We expect that the newest measurements of
$f_{gas}$ combined with non-parametric distance reconstruction from
the most recent $H(z)$ data will shed much more light on the gas
content within $R_{500}$ and its possible evolution. This paper is
organized as follows: In Section \ref{sec:daandme}, we introduce our
methodology, then we briefly describe the galaxy cluster sample from
ACTPol and the Hubble parameters from passively evolving galaxies.
The results and corresponding discussion are presented in Section
\ref{sec:anandre}. Finally, the discussion and conclusions are
respectively summarized in Section
\ref{sec:discussion}-\ref{sec:conclusion}.

\section{Methodology and data}\label{sec:daandme}

\subsection{Gas mass fraction and depletion factor}\label{subsec:fgasth}

The gas mass fraction, $f_{gas}=M_{gas}/M_{tot}$, is the ratio of
the X-ray emitting gas mass $M_{gas}$ to the cluster total mass
$M_{tot}$. Following \citet{Allen08}, the general expression for the
gas mass fraction fitted to the reference model can be given by
\begin{equation}\label{eq:fgasref}
f^{ref}_{gas}=K(z)A\gamma(z)\left(\frac{\Omega_b}{\Omega_m}\right)\left[\frac{D^{ref}_A(z)}{D_A(z)}\right]^{1.5}
\end{equation}
where $K(z)$ is the calibration constant parameterizing the
uncertainty of the instrument calibration and X-ray modeling, which
conservatively includes a 10\% Gaussian uncertainty $K=1.0\pm0.1$
\citep{Allen08}. Throughout this work a flat $\Lambda$CDM is assumed
as the fiducial cosmological model, with the matter density
$\Omega_m=0.3$, the cosmological constant representing dark energy
density $\Omega_{\Lambda}=0.7$, and the Hubble constant $H_0=70
\;\rm km s^{-1}Mpc^{-1}$. The factor $A$, which quantifies the
change in the angle subtended as the cosmology is varied, is always
very close to unity. In our analysis, we take the prior of
$\Omega_b=0.0480\pm0.0002$ and $\Omega_m=0.3156\pm0.0091$ from the
results of \cite{Planck15}. $D_A(z)$ is the true angular diameter
distance to the cluster, while $D^{ref}_A(z)$ -- its corresponding
counterpart calculated in the reference cosmology. Under the
assumption of fiducial cosmological model the latter can be
calculated as
\begin{equation}
D^{ref}_A(z)=\frac{c}{H_0}\frac{1}{1+z}\int^z_0
\frac{dz'}{\sqrt{\Omega_m(1+z')^3+(1-\Omega_m)}}
\end{equation}
Finally, the gas depletion parameter $\gamma(z)$, the ratio by which
the baryon fraction measured in clusters is depleted with respect to
the universal mean, is related to thermodynamic history of X-ray
emitting gas in the course of cluster formation. According to
Eq.~(\ref{eq:fgasref}), the observed value of $\gamma(z)$ can be
expressed as
\begin{equation}\label{eq:gammaobs}
\gamma(z)=\frac{f^{ref}_{gas}}{K}
\left(\frac{\Omega_b}{\Omega_m}\right)^{-1}
\left(\frac{D_A(z)}{D^{ref}_A(z)}\right)^{3/2}
\end{equation}
It is evident that, with a larger sample of $f_{gas}$ measurements
covering substantial redshift range, the Eq.~(\ref{eq:gammaobs}) can
provide useful insight into possible redshift evolution of the
depletion factor. Moreover, one can see that the uncertainty with
respect to the true angular diameter distances may also affect the
strength of this test. Contrary to the previous work
\citep{Holanda17a} based on luminosity distances from JLA SNe Ia
\citep{Betoule14}, translated to angular diameter distances by using
the distance duality relation $D_L(z)(1+z)^{-2}/D_A(z)=1$, in this
work we derive true angular diameter distances directly from the
cosmic chronometers $H(z)$ measurements using publicly available
code GaPP (Gaussian Process in Python) \citep{Seikel12}.

\begin{figure}[htbp]
\begin{center}
\centering
\includegraphics[angle=0,width=85mm]{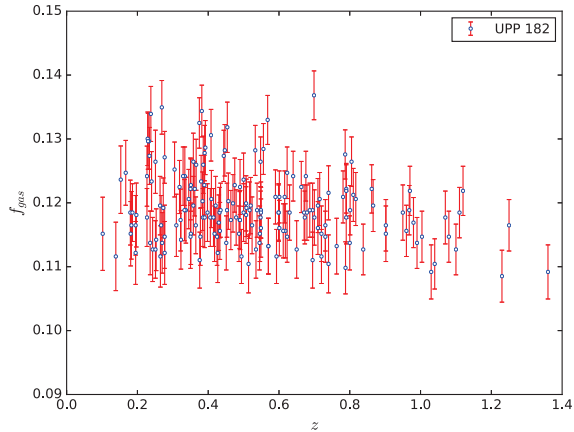}
\caption{\label{fig1} The gas mass fraction derived from ACTPol
measurements \citep{Hilton17}. Blue circles with red bars represent
central values and corresponding $1\sigma$ uncertainties calculated
from cluster total mass based on the semi-empirical relation.}
\end{center}
\end{figure}

\subsection{Galaxy cluster sample}

The gas mass fraction data used in this paper comprises 182 clusters
\citep{Hilton17} covering $0.1<z<1.4$ with the median redshift
$z=0.49$, observed by the Atacama Cosmology Telescope Polarization
experiment (ACTPol). The corresponding cluster redshift was taken
from other surveys or measured in their own follow-up observations.
In order to obtain the gas mass fraction for these galaxy clusters,
we used a semi-empirical relation verified by \citet{Vikhlinin09},
in which the gas mass fraction follows a linear relation with the
logarithm of the cluster total mass $M_{500}$ inside the radius
$R_{500}$. It should be noted that different pressure profile models
will provide slightly different total mass measurements
\citep{Hilton17}. In order to estimate this mass for each cluster,
the so-called Universal Pressure Profile (UPP) and its associated
mass-scaling relation \citep{Arnaud10} should be considered. After
the calibration of the ratio between UPP-based mass and weak-lensing
mass \citep{Planck16a, Penna17}, the final sample of $M_{500}$ data
is provided in \citet{Hilton17}. Based on these measurements, one
can assess the gas mass fraction through the following
semi-empirical relation \citep{Vikhlinin09}:
\begin{equation}\label{eq:fgasobs}
f_{gas}=0.132+0.039\log{M_{15}}
\end{equation}
where $M_{15}$ is the cluster total mass in the units of
$10^{15}h^{-1}M_{\odot}$. We remark here that, compared with the
statistical uncertainty of $M_{15}$, uncertainties of two
coefficients in the above relation are negligible \citep{Hilton17}.
Fig.~1 displays the derived $f_{gas}$ measurements and their
statistical uncertainties.

\begin{figure}[htbp]
\begin{center}
\centering
\includegraphics[angle=0,width=85mm]{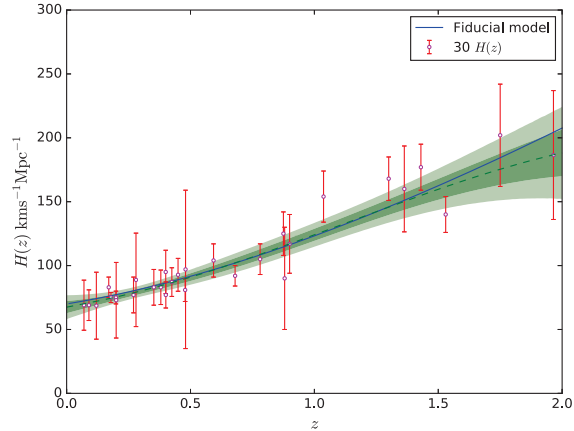}
\caption{\label{fig2} Hubble parameter measurements from cosmic
chronometers (red points) and the reconstruction of $H(z)$ function
(green envelope). Blue line corresponds to the fiducial
cosmological model. 
}
\end{center}
\end{figure}

\subsection{Cosmic chronometer sample}

Observational values of the Hubble parameter $H(z)$ at different
redshifts can be obtained through two distinct methods: cosmic
chronometers, i.e., the differential ages of passively evolving
galaxies and the BAO peak position in the radial direction. However,
as extensively discussed in the literature \citep{Zheng16},
systematic differences between these two approaches should be better
understood before one can use them jointly to get unbiased results.
Considering the fact that the only assumption for the cosmic
chronometer method is the stellar population model, which is
independent of the cosmological model, we prefer cosmic chronometers
in this paper. (See \citep{Cao11b,Cao13,Cao15} for the cosmological
applications of the Hubble parameter measurements). In particular,
following \citet{Zheng16,Zheng18,Qi18} we used the recent
compilation of 30 $H(z)$ measurements from the differential age
technique, covering the redshift range $0.07<z<1.965$ corresponding
to the redshifts of clusters for which the gas mass fraction was
measured. Using the aforementioned Gaussian processes we were able
to reconstruct the profile of $H(z)$ function up to the redshifts
$z=2$, which can subsequently be used to reconstruct the distance.
The results are shown in Fig.~\ref{fig2}, where the reconstructed
$H(z)$ function with corresponding $1\sigma$ and $2\sigma$
uncertainty strips are displayed.

\section{Analysis and Results}\label{sec:anandre}

\begin{figure}[htbp]
\begin{center}
\centering
\includegraphics[angle=0,width=85mm]{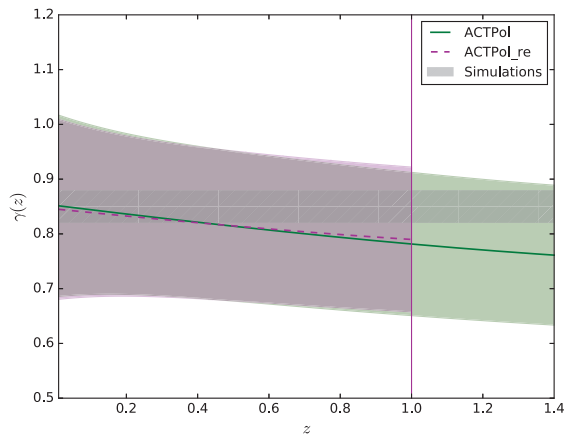}
\caption{\label{fig3} Reconstructed gas depletion factor $\gamma(z)$
for the full (green solid line) and reduced (magenta dashed line)
ACTPol cluster sample, with the shadow regions showing the 1$\sigma$
region calculated with the error propagation. The gray dashed region
corresponds to the hydrodynamical simulation results. }
\end{center}
\end{figure}

In the first step, we reconstructed the gas depletion factor as a
function of redshift. The procedure was carried out in the following
way: I) We firstly used the GPs to reconstruct $f_{gas}(z)$ function
based on the derived discrete gas mass fraction from
Eq.(\ref{eq:fgasobs}). The choice of covariance functions from the
Mant\'{e}rn family has negligible influence on the final result
\citep{Holanda17a}. II) Then the $D_A(z)$ function was reconstructed
form $H(z)$ data (assuming flat universe).
 III) Based on the reconstructed functions of
$f_{gas}(z)$ and $D_A(z)$, combined with the priors of other
relevant parameters as discussed in  Subsection~\ref{subsec:fgasth},
the $\gamma(z)$
 function  was reconstructed. Results are
shown in Fig.~\ref{fig3}.

\begin{figure}[htbp]
\begin{center}
\centering
\includegraphics[angle=0,width=80mm]{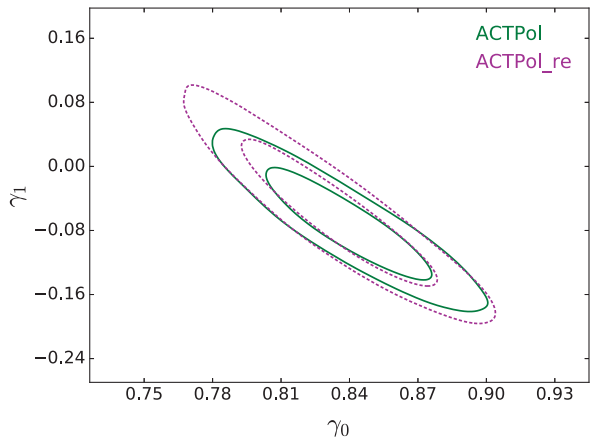}
\caption{\label{fig4} Confidence contours for the $\gamma(z)$
parameters in $\gamma(z)=\gamma_0(1+\gamma_{1}z)$. Green solid lines
and magenta dashed lines correspond to the fits obtained on the the
full and reduced ACTPol cluster sample.}
\end{center}
\end{figure}

Contrary to the previous works, the negative time evolution of the
gas depletion factor $\gamma_{500}$ can be clearly seen from the
full ACTPol cluster sample. This result is different from the
previous analysis with a smaller sample focusing on the inner region
of galaxy clusters ($r<r_{2500}$), combined with angular diameter
distances derived from the SN Ia observations. For instance, with 40
$f_{gas}$ measurements in the redshift range of $z\in[0.063,1.063]$
\citep{Allen08} and 42 $f_{gas}$ measurements in the redshift range
of $z\in[0.078,1.063]$ \citep{Mantz14}, no significant evolution of
$\gamma(z)$ was found within $R_{2500}$ \citep{Holanda17a}. In order
to compare our reconstruction with the results of hydrodynamical
simulations \citet{Battaglia13} and \citet{Planelles13} the gray
strip $\gamma=0.85\pm0.03$ is also plotted in Fig.~\ref{fig3}. One
can see that central values of our reconstructed $\gamma(z)$ are
consistent with hydrodynamical simulations up to the redshift
$z=0.4$, afterwards the reconstructed $\gamma(z)$ continues
decreasing. Not only central values but also associated $1\sigma$
strips of the reconstructed $\gamma(z)$ display decreasing trend. If
the sample of clusters was deeper than $z=1.4$ and trend did not
reverse, the strips would disconnect. Because the simulations of
\citet{Battaglia13} and \citet{Planelles13} comprised cluster
redshifts $z<1$, one might worry whether a larger redshift coverage
of the ACTPol data could be responsible for the difference in trends
seen in Fig.~\ref{fig3}. Therefore we repeated calculations on the
reduced ACTPol sample (ACTPol-re hereafter) where 10 higher redshift
(i.e. $z>1$) clusters were excluded. One can see in Fig.~\ref{fig3}
that the difference between the full and reduced ACTPol samples is
negligible. This means that the evolutionary trend $\partial
\gamma/\partial z<0$ of the depletion factor cannot be simply
attributed to the leverage of galaxy clusters located at $z>1$.
Therefore, the trend of $\gamma(z)$ decreasing with redshift
revealed in our study could reflect real evolutionary processes of
intracluster medium within $R_{500}$. Let us remark here that the
gas mass fraction obtained under hydrostatic equilibrium assumption
could be overestimated, or the true mass is underestimated
especially at large radii, which is strongly supported by recent
numerical simulations and comparisons between X-ray and lensing
masses \citep{Landry13,Giles15}. This effect could manifest itself
as lower gas depletion factor, which still needs to be investigated
with more available data.

In the second step, we investigated the issue of $\gamma(z)$
evolution using a parametric approach first used (with different
notations) in \citep{Allen08}
\begin{equation}\label{eq:gammath}
\gamma(z)=\gamma_0(1+\gamma_{1}z)
\end{equation}
where $\gamma_0$ denotes the depletion factor normalization and
$\gamma_1$ quantifies its possible evolution with redshift. Using
the Python package \textit{emcee}
\footnote{https://pypi.python.org/pypi/emcee}, which includes Markov
chain Monte Carlo (MCMC) sampler, we calculated the posterior
likelihood $\cal{L}$$\sim exp(-\chi^2/2)$, where
\begin{equation}
\chi^2=\sum^n_{i=1}\frac{(\gamma_{th}(z_i)-\gamma_{obs}(z_i))^2}{\sigma^2_{i,obs}}
\end{equation}
Theoretical expression for the depletion factor $\gamma_{th}$ and
the corresponding observational counterpart $\gamma_{obs}$ are
respectively calculated from Eq.~(\ref{eq:gammath}) and
Eq.~(\ref{eq:gammaobs}), $\sigma^2_{i,obs}$ denotes the uncertainty
of $\gamma_{obs}$ calculated according to the standard law of
uncertainty propagation using uncertainties of $K$, $\Omega_b$,
$\Omega_m$, $D_A$. Performing fits on the full and restricted ACTPol
sample, we obtained the results displayed in Fig.\ref{fig4}. The
best-fitted $\gamma$ parameters are
$\gamma_0=0.840^{+0.025}_{-0.025}(1\sigma)^{+0.048}_{-0.048}(2\sigma)$,
$\gamma_1=-0.072^{+0.044}_{-0.049}(1\sigma)^{+0.095}_{-0.086}(2\sigma)$
for the full ACTPol sample and $\gamma_0=
0.835^{+0.028}_{-0.028}(1\sigma)^{+0.056}_{-0.056}(2\sigma)$,
$\gamma_1=-0.060^{+0.056}_{-0.063}(1\sigma)^{+0.120}_{-0.110}(2\sigma)$
for the ACTPol-re sample, respectively. One can see that the fit on
$\gamma_0$ is essentially consistent with previous hydrodynamical
simulations \citep{Battaglia13, Planelles13} and observational tests
with smaller cluster sample \citep{Holanda17a, Holanda18}. However,
we also find that $\gamma_1=0$ is excluded at $1\sigma$ confidence
level for both (i.e. full and reduced) cluster samples. This is in
agreement with the reconstruction of the $\gamma_{500}(z)$ function
and indicates a mild evolutionary trend of the gas depletion factor,
which is independently supported by the angular diameter distance
measurements obtained from the SZ/X-ray technique \citep{Holanda18}.

\begin{figure}[htbp]
\begin{center}
\centering
\includegraphics[angle=0,width=80mm]{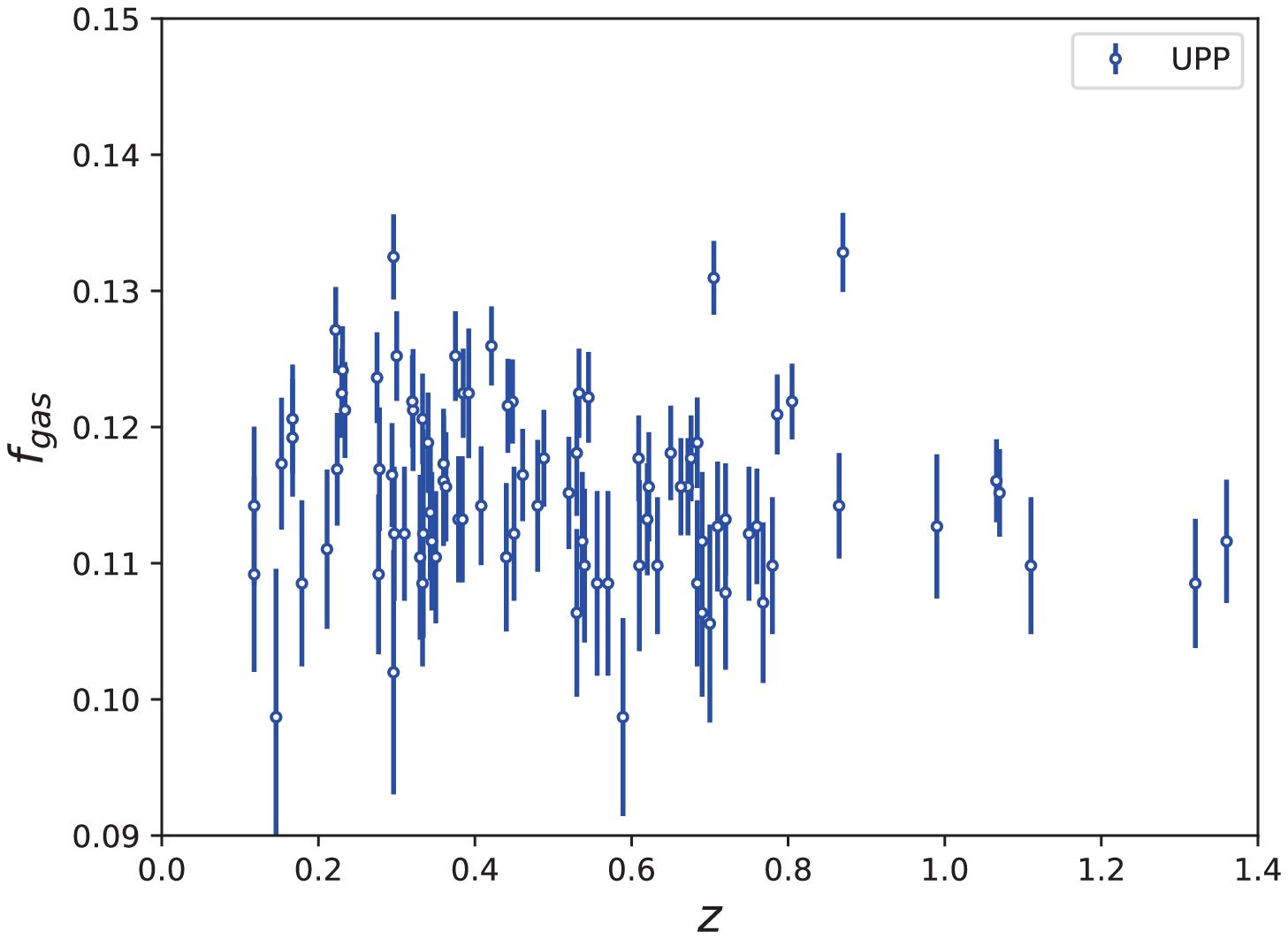}
\includegraphics[angle=0,width=80mm]{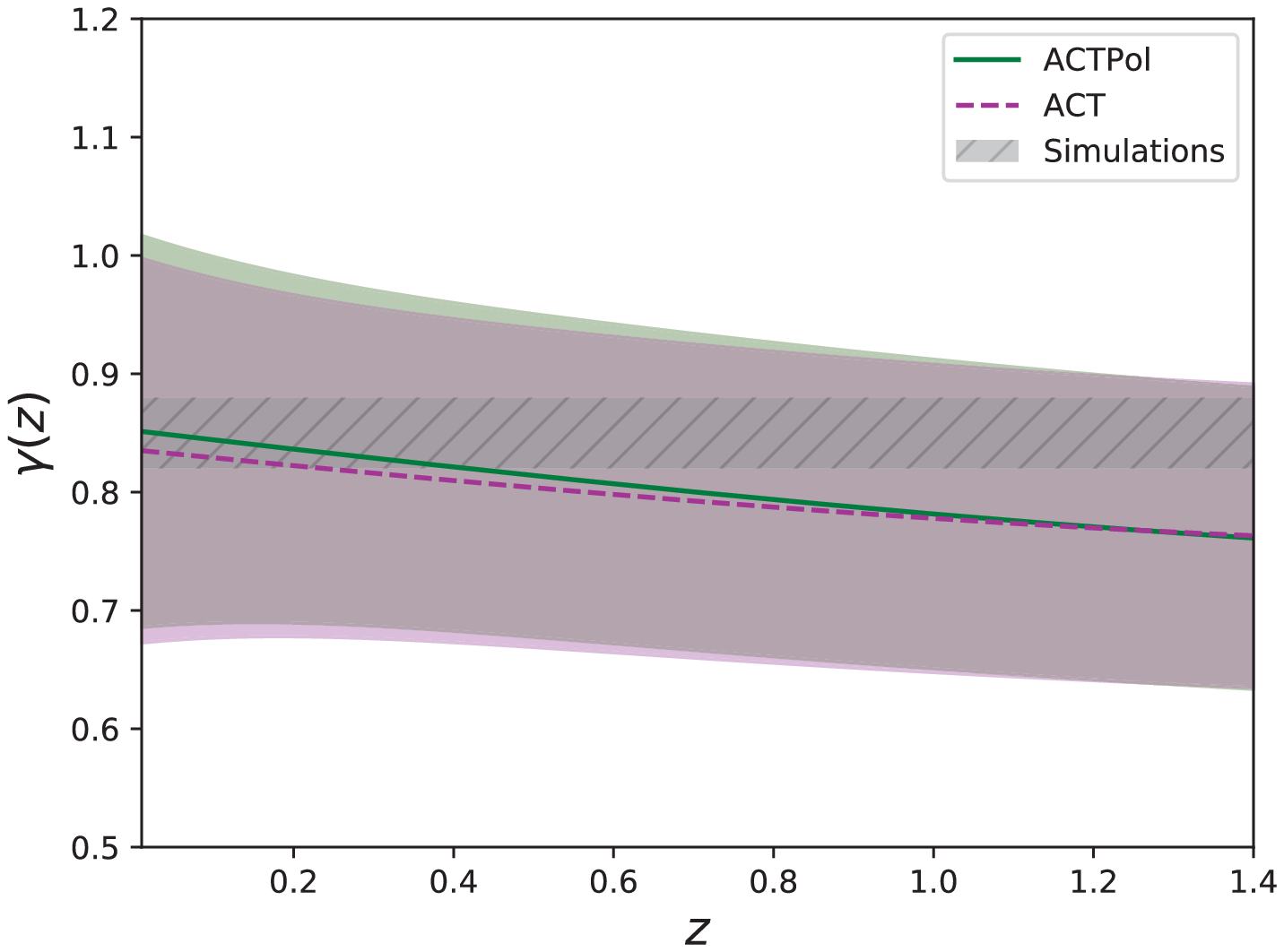}
\caption{\label{fig5} Gas mass fraction derived from ACT compilation
\citep{Hasselfield13} (left panel) and the corresponding
reconstruction of the gas depletion factor $\gamma(z)$. }
\end{center}
\end{figure}

Some sources of systematic effects that might influence our results
should be discussed. First of all, it is evident from
Eq.~(\ref{eq:gammaobs}) that $\gamma(z)$ functional dependence on
redshift is determined by the angular diameter distance ratio
$\frac{D_A(z)}{D^{ref}_A(z)}$ and possible redshift dependence of
the $f_{gas}$. However, one can see in Fig.~\ref{fig2}, that $H(z)$
function reconstructed non-parametrically from cosmic chronometers
agrees very well with the reference cosmological model. Indeed,
noticeable divergence between them can be seen at redshifts $z>1.5$
which are larger than the depth of the cluster sample. In order to
check the influence of the cosmological model on the results
quantitatively, we also repeated the calculations assuming that the
fiducial cosmological model is the true one, i.e.
$D_A(z)=D^{ref}_A(z)$. We found this influence negligible. This
means that evolutionary trend of $\gamma(z)$ discussed above could
be attributed to possible evolution of $f_{gas}$. One might raise an
objection that our result was obtained on a particular sample of
clusters and therefore could not be representative. In order to
address this issue we studied in the same way the sample of 91 SZ
detected clusters (ACT compilation) reported in
\citet{Hasselfield13} and a consistent result is given in
Fig.~\ref{fig5}. Using it we made an extrapolation to $z=1.4$ and
the question arises if it was justified. The validity or
recalibration of $M_{tot}$-$f_{gas}$ relation at higher redshifts
remains open and should be addressed in a separate study. It should
be stressed, that the gas mass fraction has been derived from
semi-empirical relation commonly used in cosmology. Such approach
was used for example, to test the validity of distance duality
relation(DDR)\citep{Goncalves15a}, the evolution of dark energy
equation of state \citep{Magana17} and the evolution of the
fine-structure constant \citep{Holanda17c}. Finally, the well-known
distance duality relation could potentially be the third important
source of systematic error on the final results. As it was
extensively discussed in the literature \citep{Holanda12}, the
equivalence of the gas mass fraction obtained from the two major
techniques, SZ effect and X-ray surface brightness observations,
might be slightly affected by the possible deviation from the
distance duality relation. Based on the $f_{gas}$ observations
derived from ACT compilation, \citet{Goncalves15a,Goncalves15b}
addressed the question of DDR on SZ gas mass fraction measurements
and concluded that the major source of uncertainty comes from the
gas mass fraction derived from SZ effect. A more detailed study of
such possibilities will be the subject of a separate paper.

\begin{table}[htp]
\begin{center}

{\begin{tabular}{l c c c c c} \hline\hline
 $f_{gas}$ sample/Distance indicator   & Cluster radius & redshift & $\gamma_0$ & $\gamma_1$ & Ref\\ \cline{1-6}
ACTPol    & $R_{500}$ & $0.1-1.4$ & $0.840^{+0.025}_{-0.025}$ & $-0.072^{+0.044}_{-0.049}$ & This work \\
ACTPol-re & $R_{500}$ & $0.1-1.0$ & $0.835^{+0.028}_{-0.028}$ & $-0.060^{+0.056}_{-0.063}$ & This work \\
\hline
 Simulation(NR) & $R_{500}$ & $0.0-1.0$ & $0.85\pm0.03$ & $0.02\pm0.05$ & \cite{Planelles13}\\
 Simulation(NR) & $R_{2500}$ & $0.0-1.0$ & $0.79\pm0.07$ & $0.07\pm0.12$ & \cite{Planelles13}\\
 \hline
 $f_{gas}$/SN Ia & $R_{2500}$ & $0.078-1.063$ & $0.85\pm0.08$ & $0.00\pm0.05$ & \cite{Holanda17a}\\
$f_{gas}$/Cluster I & $R_{2500}$ & $0.14-0.89$ & $0.76\pm0.14$ & $-0.42^{+0.42}_{-0.40}$ & \cite{Holanda18}\\
$f_{gas}$/Cluster II & $R_{2500}$ & $0.14-0.89$ & $0.72\pm0.01$ & $0.16\pm0.36$ & \cite{Holanda18}\\
$f_{gas}$/$\Lambda$CDM & $R_{2500}$ & $0.12-0.78$ & $0.84\pm0.07$ & $-0.02\pm0.14$ & \cite{Holanda18}\\
 \hline \hline
\end{tabular} \label{table}}
\caption{Summary of the best-fitted gas depletion factor parameters
and the corresponding $1\sigma$ uncertainty in this analysis and in
the literature \citep{Planelles13,Holanda17a,Holanda18}, obtained at
different cluster radius ($R_{500}$ and $R_{2500}$) from
non-radiative(NR) simulations \citep{Planelles13} and different
$f_{gas}$ samples \cite{LaRoque06,Hilton17}. In the previous works,
the luminosity distances/angular diameter distances are derived from
recent SN Ia observations, SZ effect/X-ray measurements of Cluster I
\citep{Bonamente06} and Cluster II \citep{Filippis05}, or in the
framework of Planck's best-fitted $\Lambda$CDM cosmology
\citep{Planck15}.}

\end{center}
\end{table}

\begin{figure}[htbp]
\begin{center}
\centering
\includegraphics[angle=0,width=85mm]{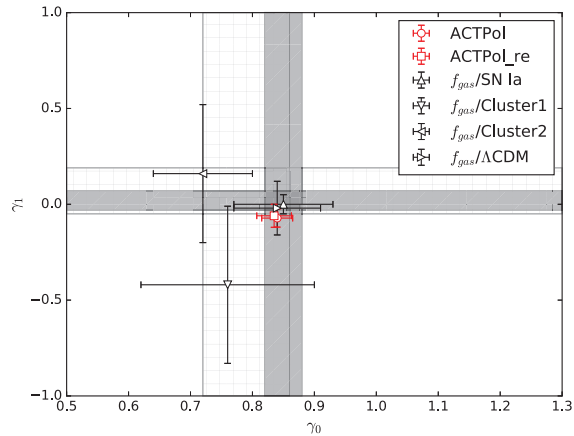}
\caption{\label{fig6} Comparison of the best-fitted gas depletion
factor parameters ($\gamma_0, \gamma_1$). The red circle and square
denote the best-fitted gas depletion factor within the cluster
radius of $R_{500}$, concerning the whole ACTPol and reduced ACTPol
sample, respectively. The grey dashed and white netted region show
the hydrodynamical simulation results ($1\sigma$ uncertainties) at
different cluster radius ($R_{500}$ and $R_{2500}$), while the
triangles represent the best-fitted gas depletion factor parameters
within $R_{2500}$, given the same $f_{gas}$ sample and different
distance indicators.}
\end{center}
\end{figure}

\section{Discussion}\label{sec:discussion}

Now one important issue is the comparison of our results with those
of earlier studies done using other, alternative methodologies. The
numerical results are summarized in Table~I. Using a set of
hydrodynamical simulations of galaxy clusters characterized by
different physical processes, \citet{Planelles13} explored how the
fraction and spatial distribution of baryons (contributed both by
the stellar component and the hot X-ray emitting gas), are affected
by the feedback from supernova (SN) and active galactic nuclei (AGN)
within $R_{500}$ and $R_{2500}$. More specifically, the depletion
factor $\gamma(z)$, as well as its dependence on redshift, baryonic
physics, and cluster radius were thoroughly discussed. Their results
showed that the depletion in baryon content within $R_{500}$ is more
pronounced, with a stronger mass dependence for the simulations
including AGN feedback. However, in the framework of the functional
form of $\gamma(z)=\gamma_0+\gamma_1z$, the baryon depletion factor
does not evolve significantly with redshift ($z<1$), regardless of
the considered radius or physics. At last, the simulation results
gave the best-fit parameter: $\gamma_0=0.85\pm0.03$,
$\gamma_1=0.02\pm0.05$ at $R_{500}$ and $\gamma_0=0.79\pm0.07$,
$\gamma_1=0.07\pm0.12$ at $R_{2500}$ \citep{Planelles13}. On the
other hand, the attempt to determine the baryon depletion factor
with currently available observations was presented in
\citet{Holanda17a}, which investigated the viability of using 40
X-ray emitting gas mass fraction measurements \citep{LaRoque06} and
luminosity distance measurements from SNe Ia \citep{Betoule14}
(based on the validity of distance duality relation \citep{Cao11c})
to place additional constraints on the behavior of $\gamma(z)$. It
was found that $\gamma_0=0.85\pm0.08$ and $\gamma_1=0.00\pm0.05$
within $R_{2500}$, from which one may observe the well consistency
between fits obtained from current observations and hydrodynamical
simulations. Further papers have also studied the possible time
evolution for $\gamma(z)$, in light of exclusively galaxy cluster
data. More recently, Ref.~\cite{Holanda18} proposed a new method to
investigate the depletion factor, in light of X-ray gas mass
fraction and angular diameter distance measurements from
Sunyaev-Zel'dovich effect plus X-ray observations
\citep{Bonamente06,Filippis05}. Note that in their analysis, the
electron density and temperature profiles of galaxy clusters, which
provide the measurements of angular diameter distances, are
described by the non-isothermal double $\beta$-model or under the
assumptions of spherical symmetry and hydrostatic equilibrium
\citep{Filippis05,Bonamente06}. Different from the findings of the
simulations, the analysis has revealed a non-negligible time
evolution for the depletion factor: $\gamma_1=-0.42^{+0.42}_{-0.40}$
and $\gamma_1=0.16\pm0.36$. Such conclusion, however, disagrees with
the constraints on depletion factor ($\gamma_0=0.84\pm0.07$,
$\gamma_1=-0.02\pm0.14$) by using the same $f_{gas}$ sample and
angular diameter distances obtained from the flat $\Lambda$CDM model
(Planck results) \citep{Holanda18}. Therefore, the importance of
non-parametric reconstruction of angular diameter distance using
Hubble parameters from cosmic chronometers are indeed revealed in
this analysis.

We also provide a graphical representation of the comparison results
in Fig.~\ref{fig6}, which directly shows the depletion factor
parameters obtained in this analysis and the previous works (see
Table I for details). The red circle and square denote the
best-fitted gas depletion factor within the cluster radius of
$R_{500}$, concerning the whole ACTPol and reduced ACTPol sample,
respectively. The grey dashed and white netted region show the
hydrodynamical simulation results ($1\sigma$ uncertainties) at
different cluster radius ($R_{500}$ and $R_{2500}$), while the
triangles with different directions represent the best-fitted gas
depletion factor parameters within $R_{2500}$, concerning the
\citet{LaRoque06} $f_{gas}$ sample and different distance indicators
(SN Ia observations, SZ effect/X-ray measurements of galaxy
clusters, and Planck's best-fitted $\Lambda$CDM cosmology). On the
one hand, we find that the $\gamma_0$ value is in full agreement
with the simulated results derived within $R_{500}$. On the other
hand, although the $\gamma_1$ value in our analysis is compatible
with $\gamma_1=0$ within 2$\sigma$, a non-negligible time evolution
for the depletion factor is still supported by the current
observations. Such tendency is clearly in tension with the results
of cosmological hydrodynamical simulations \citep{Planelles13}, but
well consistent with the self-consistent observational constraints
by using exclusively galaxy cluster data \citep{Holanda18}.

\section{Conclusions}\label{sec:conclusion}

We studied the evolution of the gas depletion factor $\gamma(z)$
inside the radius $R_{500}$, using the largest SZ cluster sample
obtained by the Atacama Cosmology Telescope Polarization experiment
(ACTPol) \citep{Hilton17}. The sample comprised 182 galaxy clusters
covering the redshift range of $0.1<z<1.4$. Using two methods:
non-parametric reconstruction of $\gamma(z)$ and fitting $\gamma_0,
\gamma_1$ parameters in the evolutionary model
$\gamma(z)=\gamma_0(1+\gamma_{1}z)$ we revealed an unambiguous trend
of $\gamma(z)$ decreasing with redshift. This is contrary to recent
claims of \citet{Holanda17a} who found no evidence for such
evolution. However, their analysis was focused on $\gamma(z)$ inside
the inner region of galaxy clusters ($r<r_{2500}$) and performed on
a smaller sample. It should be noted that the best-fitted value our
reconstructed $\gamma(z)$ is well consistent with the hydrodynamical
simulations at $z<0.4$ \citep{Battaglia13,Planelles13}. However,
when the 1$\sigma$ uncertainty is taken into account, the
$\gamma(z)$ reconstructed from the full ACTPol sample and
hydrodynamical simulations overlap with each other. On the other
hand, the reconstructed uncertainty strip displays an unambiguous
trend while the simulation results stay constant. They would
eventually detach once we had access to cluster data at higher
redshifts. Moreover, parametric fits excluded no evolution case of
$\gamma_1=0$ at $1\sigma$ confidence level. These results have been
confirmed on the reduced ACTPol measurements, the redshift coverage
of which is consistent with that of hydrodynamical simulations
($z<1$) and on the alternative sample of 91 SZ clusters reported
earlier in ACT compilation. Discussion of possible systematic
effects leaves one open question about validity of the empirical
relation $M_{tot}$-$f_{gas}$ obtained on very close clusters.

Summarizing, the results presented in this paper could pave the way
to explore the hot gas fraction within large radii of galaxy
clusters as well as its possible evolution with redshift, which
should be studied further on larger galaxy cluster samples available
in the upcoming X-ray/SZ cluster surveys. With the dawn of the era
of GW astronomy, which was opened by the first direct detection of
gravitational waves (GWs) by the LIGO/Virgo collaboration
\citep{Abbott16}, one could expect the possibility of testing
$\gamma(z)$ at much higher precision in the future, along with the
observational search for more GW events with smaller statistical and
systematic uncertainties \citep{Cai17,Qi19a,Qi19b}.

\acknowledgments{This work was supported by  the National Key
Research and Development Program of China under Grants No.
2017YFA0402603; the National Natural Science Foundation of China
under Grants Nos. 11690023 and 11633001; the Beijing Talents Fund of
Organization Department of Beijing Municipal Committee of the CPC;
the Strategic Priority Research Program of the Chinese Academy of
Sciences, Grant No. XDB23000000; the Interdiscipline Research Funds
of Beijing Normal University; and the Opening Project of Key
Laboratory of Computational Astrophysics, National Astronomical
Observatories, Chinese Academy of Sciences. J.-Z.Q. was supported by
the China Postdoctoral Science Foundation under grant No.
2017M620661. M.B. was supported by the Foreign Talent Introducing
Project and Special Fund Support of Foreign Knowledge Introducing
Project in China.}

\end{document}